# FPGA Implementation of ECG feature extraction using Time domain analysis

*Naveen Sai Madiraju*[1], *Naresh Kurella*[2], *Rama Valapudasu*[3]

Department of Electronics and Communication Engineering
National Institute of Technology, Warangal
Warangal, India

E-mail: naveensai.madiraju@gmail.com[1], naresh.kurella@gmail.com[2], vr.nitw@gmail.com[3]

*Abstract*— An electrocardiogram (ECG) feature extraction system has been developed and evaluated using Virtex-6 FPGA kit which belongs to Xilinx Ltd. In time domain, Pan-Tompkins algorithm is used for QRS detection and it is followed by a feature extractor block to extract ECG features. This whole system can be used to detect cardiac arrhythmia. The completed algorithm was implemented on Virtex-6(XC6VLX240-T) device and tested using hardware co-simulation in Modelsim and simulink environment. The software generated ECG signals are obtained from MIT-BIH arrhythmia Database [1]. The memory and time complexities of the implemented design were recorded and feature extraction has been done. We have achieved satisfactory results which is mainly due to parallel implementation. Therefore accurate arrhythmia detection using hardware implementation a viable approach.

## I. INTRODUCTION

ECG is an acronym for Electrocardiogram which gives detail about the electrical activity of heart. Most common cardiac abnormalities can be identified using ECG and heart rate analysis. Hence the study of ECG signals is of great importance for arrhythmia understanding. A statistical study of the extracted features indicate a significant deviation between different arrhythmia types and normal heart . Therefore ECG signals feature extraction and statistical analysis will be of immense importance in arrhythmia identification.

Many ECG feature extraction algorithms and techniques have been proposed in the literature[2][3][4]. A few algorithms have been proposed based on first derivative and amplitude [5][6][7] . Many of these techniques are focused on extraction of QRS-complex. Kohler et al.[8] carried a comprehensive study on various QRS detection methods. Subsequently Pan-tomkin [9] developed an algorithm exploiting the steep slope of QRS complex for feature extraction. Later Laguna et. al.[10], in multi lead ECG systems made use of a Pan-Tompkins based algorithm to estimate ECG wave boundaries. More robust estimation has been made by combining results from different leads. Lately Wavelet Transform has been extensively used for ECG signal feature extraction[11][2][12]. In wavelet analysis ECG signal is transformed into frequency domain corresponding to their time localizations. ECG features can be extracted by analyzing Wavelet Transform coefficients at different decomposition levels. An acceptable tolerance limit for the ECG feature extraction algorithms has been provided by "Common Standards for Quantitative Electrocardiography" (CSE)[13].

A real time accurate ECG feature extractor would be greatly useful in field of medicine. Many technologies have been developed lately to solve the problem of real time data acquisition, processing and transmission. Recently reconfigurable systems like Field Programmable Gated Array (FPGAs) have become popular due to massive real time parallel processing capability. However, FPGA implementation of ECG feature extraction algorithm becomes a difficult problem due to time varying nature of the signal when subjected to stress and due to the presence of noise. Recently pavlatos et al.[3] made a detailed study on FPGA implementation of pan-Tomkin algorithm, which made use of a non pipelined architecture making it unable to run in real time . Later Nambakhsh et al.[2] proposed a FPGA based feature extractor and classifier making use of pipelined FPGA architecture in wavelet domain.

In this study we propose a real time pipelined pan-tomkin based feature extractor. Our proposed method performs feature extraction within FPGA's framework. To do this we make use of pan-tomkin algorithm for QRS complex identification, further we extract R-R peak interval, width of QRS complex and Heart rate features.

The rest of the paper is structured as follows. The subsequent section gives the architecture of pan-tomkin algorithm and feature extractor , while the experimentation and results are dealt in next chapter. The final section concludes this study.

## II. PROPOSED MODEL

The proposed model consists of two main components. The first component is the QRS detector. We have used pan-tomkin algorithm for QRS detection. The second component is Feature extractor. Fig.1 displays the flow diagram of proposed method. In the subsequent discussion we give detail presentation the individual blocks.

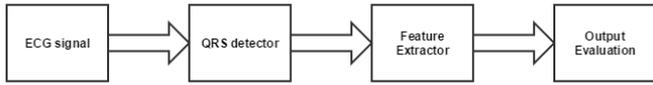

Fig. 1. Block diagram of proposed model

## A. The QRS detector

First, Pan Tomkin algorithm is indeed one of the most popular QRS detection technique used virtually in all biomedical signal processing applications. Figure 2 shows an overview of the algorithm.

1. First the raw ECG signal is passed through a low pass filter and a high pass filter to remove baseline wander and power line interface and noise. (output denoted by SF). The low pass filters can be formulated as:

$$H(z) = \frac{(1-z^{-6})^2}{(1-z^{-1})^2} \quad (1)$$

$$y(n) = 2y(n-1) - y(n-2) + x(n) - 2x(n-6) + x(n-12) \quad (2)$$

The high pass filter can be formulated as:

$$H(z) = \frac{1}{32} \frac{(1-z^{-32})}{(1-z^{-1})} \quad (3)$$

$$y(n) = y(n-1) - \frac{1}{32}x(n) - x(n-16) - x(n-17) + \frac{1}{32}x(n-32) \quad (4)$$

2. After noise removal, the signal is passed through a differentiator. The differentiator can be mathematically modelled as:

$$H(z) = \frac{1}{10}(2 + z^{-1} - z^{-3} - 2z^{-4}) \quad (5)$$

$$y(n) = \frac{1}{8}[2x(n) - x(n-1) - x(n-3) + 2x(n-4)] \quad (6)$$

3. After differentiation squaring is done point by point to make all the data positive. Squaring can be mathematically formulated as:

$$y(n) = [x(n)^2] \quad (7)$$

4. After squaring, a window 150 ms is chosen and sliding window integration is done to extract the QRS complex. (output denoted by SI). Mathematically moving window integration is described as follows:

$$y(n) = \frac{1}{32}\sum_{i=1}^{32} x(n-i) \quad (8)$$

A threshold T is calculated using moving mean of SI and SF having window length of 150 ms respectively. The threshold T is a linear combination of signal peak and noise peak as used in pan tomkin algorithm [9]. The signals are forwarded to feature extractor if both of SF and SI crossed their respective thresholds.

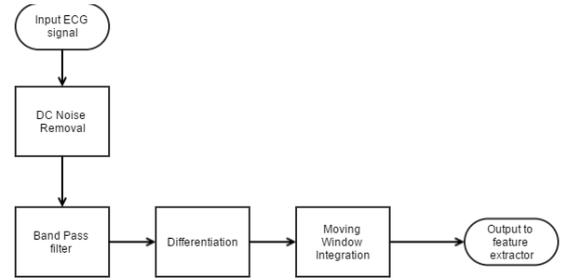

Fig. 2. Flow of QRS detection algorithm

## B. The feature extractor

The output of the sliding window integrator and the filtered ECG signal is given to the feature extractor. The following features were extracted for subsequent ECG classification:

1. The QRS width was extracted from the output of Sliding window integrator. The rise time of the output of pan-tomkin algorithm is the QRS width. Hence, for calculation rise time we implement a combinational circuitry to count till the input rises. An automatically adjustable threshold T is used to trigger the counter. After reaching peak a self delay of 100 ms is introduced to avoid false outputs and the QRS width is extracted. The Pseudo code of QRS interval extractor is shown in Alg. 1.

```
Data: Output of Moving window integrator
Result: QRS interval
initialization;
while 1 do
    read current;
    if y > y/z AND y > T then
        start count;
        if R-peak reached then
            QRS-interval=count;
            Reset count;
            Delay 100 ms;
        end
    end
end
```
Algorithm 1: QRS interval extraction

2. R peaks were identified and R-R peak interval is extracted. To identify R-peaks we realize the fact that local maxima are the position of R-peak, tracked using first derivative. The time interval difference between two subsequent R-peaks will give R-R peak interval feature. The Pseudo code of R-R interval extractor is shown in Alg. 2.

```
Data: Output of Moving window integrator
Result: R-R interval
p(n) = y(n) − y(n − 1) initialization;
while 1 do
    read current;
    if p ≤ 0 AND NOT p/z ≤ 0 then
        if past 50 samples ≥ 0 then
            Rpresent=countval;
            R-R interval=Rpresent-Rprev;
            Rprev=Rpresent;
        end
    end
end
```
    **Algorithm 2:** R-R interval extraction

3. Furthermore, Heart Rate (HR) is also extracted using the output from QRS detector. HR can be evaluated using following equation:

$$HR = 60 \frac{\text{sampling frequncy}}{\text{R-R interval}} \quad (9)$$

### III. ARCHITECTRE OF DESIGN

The implementation of the algorithm can be done in various architectural styles: parallel, semi-parallel and sequential. In this paper we have used parallel architecture for FPGA implementation to ensure real time classification. The model based design will ensure error free fast FPGA prototyping and design. Many tools have been provided by MATLAB which automates FPGA design and enable validation in the form of software and hardware co-simulation. Figure 3 shows simulink model based design of the algorithm. We have used the difference equations formulated in pan tomkin algorithm to design the simulink model.

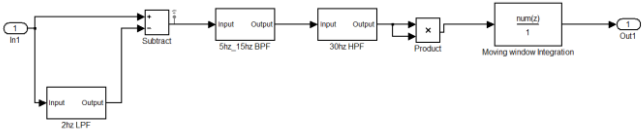

Fig. 3. Simulink model of QRS detector

All the processing is done with fixed point arithmetic so that the algorithm can operate in real time without requiring excessive computing power. We have used fixed point tool provided by matlab to know most suitable exponent and fraction length. The following Fig. 4 shows the RTL schematic of the algorithm. A parallel architecture based model has been designed. In this paper Xilinx virtex-6(XC6VLX240-T) development board has been used.

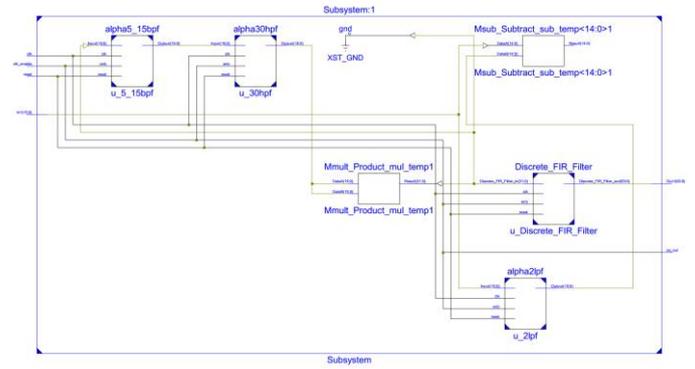

Fig. 4. RTL schematic of QRS detector

In FPGA inputs and outputs are in the form of bit streams, the visualization of the output becomes extremely difficult. Simulink environment provides option for hardware co-simulation, JTAG is used to realize communication between simulink and FPGA. The ECG signal input in simulink environment is converted into bit stream, through JTAG cable send to FPGA, FPGA runs the algorithm, output is again received to the simulink environment through the JTAG cable. There is a provision of comparing output of simulink model and FPGA model. In this paper we have achieved zero deviation between FPGA model and simulnk model.

### IV. EXPERIMENTATION AND RESULTS

#### A. Qualitative analysis

To evaluate the system performance we used MIT-BIH arrhythmia database. To visualize the results of hardware co-simulation we make use of modelsim PE student edition and simulink environment. Fig.5 shows the output of the moving window integrator in bit stream form in modelsim. Since is difficult to examine the results as bit stream, we make use of simulink environment to display the results. Fig.6 shows the hardware co-simulation output at each stage of the algorithm. In the subsequent section we would discuss the results obtained.

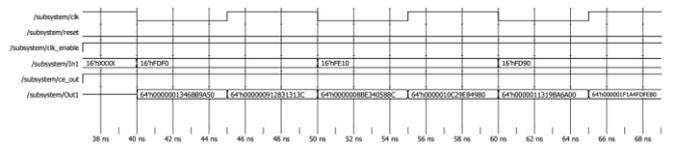

Fig. 5. Inputs and Output of Pan-Tomkin FPGA model in modelsim using co-simulation.

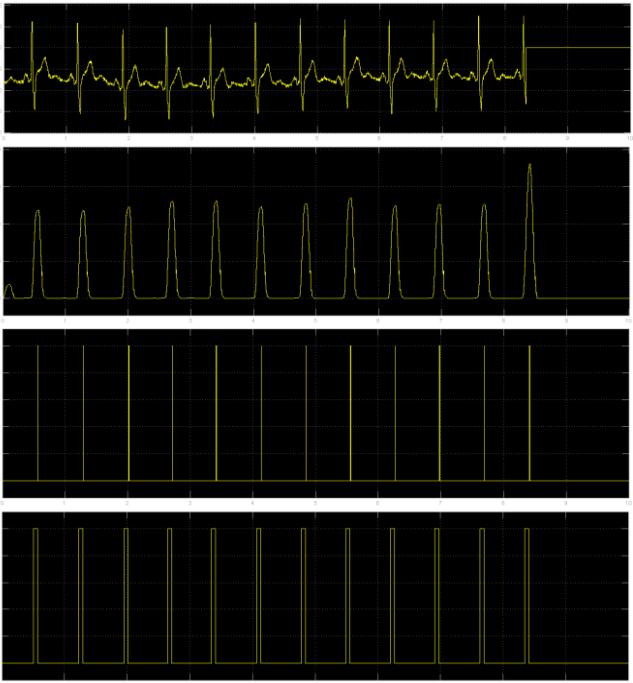

Fig. 6. Application of proposed algorithm to ECG signal. Row 1: ECG signal. Row 2: Output of Pan-Tomkin algorithm. Row 3: Extracted R-peaks. Row 4: Extracted QRS Interval.

*B. Quantitaive analysis*

The hardware resource utilization of virtex-6 FPGA is shown in Table I. The figures in Table I reveal the availability of enough resources for additional functionalities. A report on timing analysis is presented in Table II showing small combinational delay enabling the algorithm to run in real time.

TABLE I. HARDWARE RESOURCE UTILIZATION

| *Logic utilization* | *used* | *available* | *utilization* |
|---|---|---|---|
| Slice registers | 1540 | 93120 | 1% |
| Slice LUTS | 4324 | 46560 | 9% |
| LUT-FF pairs | 260 | 5604 | 4% |
| DSP48E1 | 125 | 288 | 43% |
| BUFG/BUFGCTRLs | 1 | 32 | 3% |
| Bonded IOBs | 30 | 240 | 12.5% |

TABLE II. TIMING REPORT

| | |
|---|---|
| Minimum period | 67.74 ns |
| Maximum frequency | 14.762 MHz |
| Minimum input arrival time before clock | 68.820 ns |
| Maximum output required time after clock | 107.306 ns |
| Maximum combinational path delay | 108.386 ns |

## V. CONCLUSION AND FUTURE WORK

The FPGA implementation of effective algorithm for fast ECG feature extraction would be helpful in making efficient devices having ECG analysis capability. Parallel architectures can be realized using FPGA's. The proposed algorithm has proven its low resource utilization and accurate feature extraction. In the near future wearable devices with real time ECG analysis capability can make use of this architecture.

Our future work is aimed at developing a FPGA prototype of higher order spectral analysis based ECG feature extractor and classifier. We would like to include a multilayer perceptron in the classifier for accurate cardiac arrhythmia detection.